\documentclass[a4paper,11pt]{article}
\pdfoutput=1 
\usepackage{jinstpub} 
\usepackage{graphicx}
\usepackage{subfigure}
\usepackage{siunitx}
\usepackage{multirow}
\pdfoutput=1
\usepackage{pdfpages}
\DeclareGraphicsExtensions{.jpeg}

\title{\boldmath Single-Electron Spectra in RPWELL-based detectors}

\author[a,1,2]{Purba Bhattacharya,\note{Corresponding author.}\note{Presently at Dept. of Physics, University of Calcutta, India.}}
\author[a]{Andrea Tesi,}
\author[a]{Dan Shaked-Renous,}
\author[b]{Luca Moleri,}
\author[a]{Amos Breskin,}
\author[a]{Shikma Bressler.}

\affiliation[a]{Department of Particle Physics and Astrophysics, Weizmann Institute of Science, 7610001 Rehovot, Israel}
\affiliation[b]{Technion - Israel Institute of Technology, 3200003 Haifa, Israel}

\emailAdd{purba.bhattacharya85@gmail.com}

\abstract{Single-electron spectra are the key ingredient in the efficient 
detection of single UV-photons. 
In this work, we investigated the shape of single-photoelectron spectra in 
single- and double-stage Resistive Plate WELL (RPWELL) detector 
configurations, operated in $\mathrm{Ne/CH_{4}}$ and $\mathrm{Ar/CH_{4}}$. 
Discharge-free operation was reached over a broad dynamic range, with charge 
gains of \numrange[range-phrase = -]{e4}{e6}. 
Compared to the usual exponential ones, the observed Polya-like charge spectra 
pave the way towards higher single-electron detection efficiencies. 
The latter was evaluated here, using experimental data combined with numerical 
simulations.
The effects of the gas mixtures, electric field configuration, and detector 
geometry on the Polya spectra and their related ``$\theta$'' parameter are 
presented.}

\keywords{Charge transport and multiplication in gas; Electron multipliers (gas); Micropattern 
gaseous detectors (MSGC, GEM, THGEM, RETHGEM, MHSP, MICROPIC, MICROMEGAS, InGrid, etc), Photon 
Detector for UV, visible and IR photons (gas) (gas-photocathodes, solid-photocathodes)}

\begin{document}
\maketitle
\flushbottom


\sisetup{range-phrase=--}

\section{Introduction}

Gas-avalanche detectors, introduced originally for the needs of particle-physics experiments, have 
become a prominent subject of research also in a variety of other fields \cite{Ref1}. 
To this extent, significant efforts have been dedicated to the development of single-electron 
\cite{Ref2, Ref3} and single-photoelectron UV-photon gaseous detectors; the latter, mainly in the 
context of Ring Imaging Cherenkov Counters (RICH) \cite{Ref4, Ref5}, with gaseous and solid 
photocathodes. 
Such gaseous photomultipliers (GPM) \cite{Ref6, Ref7} provide a cost-effective solution suitable 
for the coverage of large areas with good spatial and temporal resolution, and low sensitivity to 
magnetic fields. 
GPM detectors have been developed also for imaging scintillation and electroluminescence photons in 
noble gases \cite{Ref8} and liquids \cite{Ref9}.

In this context, Gaseous Photodetectors \cite{Ref6} employing Multiwire \cite{Ref10} and Drift 
\cite{Ref11} Chambers, Multi-step avalanche chambers \cite{Ref12}, cascaded GEMs 
\cite{Ref13, Ref14, Ref15} and THGEM-based \cite{Ref16, Ref17, Ref18, Ref19} detectors have been 
playing an important role in experiments. 
As demonstrated in \cite{Ref20}, the more recent ``hybrid'' CsI-coated THGEM-Micromegas has shown to 
be an efficient upgrade for the COMPASS-RICH-I. 
Furthermore,  THGEM-based detectors were demonstrated to have moderate (sub-millimeter) localization
 resolution \cite{Ref21} and about $\mathrm{10~nsec}$ time resolution, which comply with the 
requirements of most RICH devices. 
Though, previous studies with THGEM detectors have shown high gas gains for single-photoelectron 
detection \cite{Ref17}, the latter could be considerably lower in the presence of intense particle 
background \cite{Ref22}.

\begin{figure}[hbt]
\centering
\subfigure[]
{\label{Single}\includegraphics[scale=25]{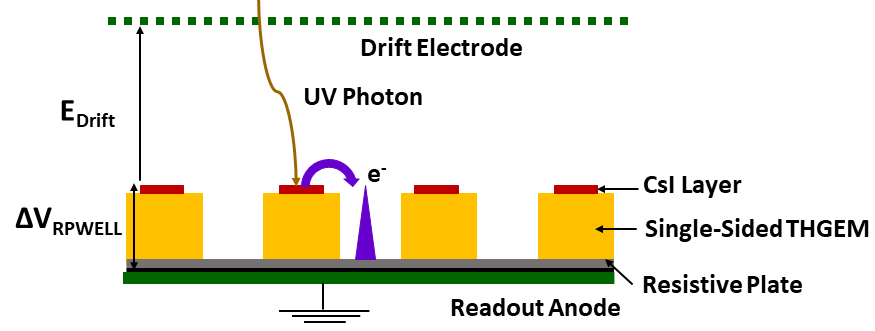}}
\subfigure[]
{\label{Double}\includegraphics[scale=50]{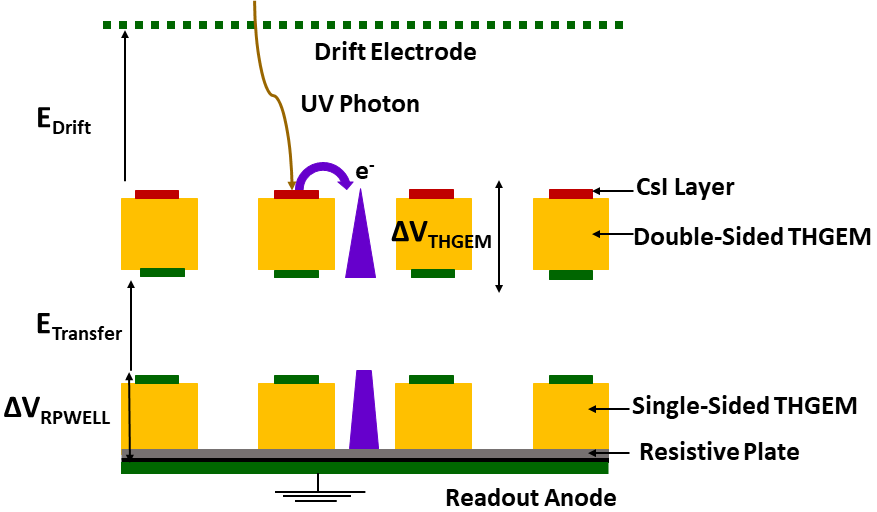}}
\caption{A schematic view of the (a) single-stage RPWELL detector and (b) double-stage RPWELL-based
detector with a double-sided THGEM pre-amplification stage.}
\label{Detector}
\end{figure}

In some applications, highly ionizing background environment can result in the formation of large 
avalanches, often leading to electric discharge. 
The latter can damage the readout electronics and the detector's electrodes; it often introduces 
significant dead-time. 
The thick Resistive WELL (RWELL) \cite{Ref22} and the Resistive Plate WELL (RPWELL) detectors 
\cite{Ref23} were introduced to prevent occasional discharges and mitigate their potential 
destructive effects. 
In the RPWELL (Fig.~\ref{Single}), the single-sided CsI-coated THGEM electrode is coupled to a 
readout anode through a thin plate of high bulk resistivity 
(\SIrange[range-units=single]{e9}{e12}{~\mathrm{\Omega cm}}). 
Ionization electrons induced by X-rays or UV-induced photoelectrons from a photocathode deposited 
on the top surface (e.g. CsI) are collected into the WELL holes where they undergo avalanche 
multiplication. 
Signals are induced capacitively through the resistive plate onto a patterned readout anode, in 
direct contact with the resistive plate. 
The RPWELL and its properties have been studied extensively in the laboratory and with particle 
beams \cite{Ref24, Ref25, Ref26}. 
The studies demonstrated a high particle detection efficiency, over a broad dynamic range and at 
high particle-flux range, in a discharge-free operation mode at charge gains up to ${\sim}10^{4}$. 
Compared to single-stage configurations, higher maximal achievable gains and lower discharge 
probabilities can be achieved in a multi-stage one. 
In the present work, a double-sided THGEM followed by a RPWELL was investigated (Fig.~\ref{Double}). 
This leads to higher detector gains at lower voltage bias per single THGEM and RPWELL element and 
thus to higher operation stability.

The single-photon detection efficiency (PDE) of a gaseous photon detector with a photocathode 
followed by an amplification element is provided by:

\begin{equation}
\label{eqn1}
\it{\epsilon_{eff}(\lambda)} = \it{QE(\lambda)\epsilon_{extr}\epsilon_{coll}\epsilon_{thresh}}
\end{equation}

Here, $QE(\lambda)$ is the wavelength-dependent quantum efficiency value in vacuum of the 
photocathode. 
$\epsilon_{extr}$ is the photoelectron extraction efficiency into the gas \cite{Ref27, Ref28}, 
$\epsilon_{coll}$ is the efficiency to transfer the extracted photoelectron into the amplification 
region \cite{Ref17}. 
$\epsilon_{thresh}$ is the single-electron detection efficiency - the probability that the 
collected photoelectron will generate a signal above a given threshold. 
$\epsilon_{thresh}$ is strongly related to the shape of the electron charge spectrum which reflects the 
physical processes governing the formation and development of the single-electron avalanche.

In the present study, we investigated the UV-induced single-photoelectron charge spectra generated 
in $\mathrm{Ne/CH_{4}}$ and $\mathrm{Ar/CH_{4}}$ gas mixtures with CsI-coated single RPWELL and 
double-stage CsI-coated THGEM followed by an RPWELL. 
In section 2 we elaborate on the theory associated with the formation of single-electron charge 
spectra. 
The experimental setup is detailed in section 3, followed by the introduction of the numerical 
analysis framework in section 4. 
The results are presented in section 5 and discussed in section 6.

\section{Single-electron charge spectra in gas-avalanche detectors - theoretical aspects}

Electron avalanches in gas develop stochastically; the mean-free-path between successive 
interactions and, accordingly, the electron energy available, can vary considerably between 
successive interactions. 
This is dictated by the cross sections of the different electron interactions with gas molecules 
(elastic and inelastic). 
As a result, the avalanche size (e.g. detector charge gain) fluctuations determine the shape
 of single-photoelectron charge spectra (see detailed study by Alkhazov \cite{Ref29}). 
Thus, these statistical fluctuations set a physical limit on the single-photoelectron detection 
efficiency and localization resolution. 
Furthermore, large avalanche fluctuations increase the probability for high-charge events, with a 
potential for occasional discharges. 

The shape of the single-electron charge distribution depends on the gas mixture, the reduced 
electric field ($\it{E/p}$; $\it{p}$ being the gas pressure and $\it{E}$ the electric field), the 
electron initial momentum and the distance over which the avalanche develops. 
Gain fluctuations can be described quantitatively in terms of the probability 
$\it{P_{n}(r_{0}, p_{0})}$ that an electron with initial momentum $\it{p_0}$ released at a position 
$\it{r_0}$ initiates an avalanche resulting in $\it{n}$ electrons in the detector. 

According to \cite{Ref30, Ref31}, under a moderate uniform field, an estimate of the single-electron
 avalanche distribution can be carried out with the assumption that the probability of ionization by
 an electron depends only on the electric field strength and is independent of its previous history.
 Yule-Furry statistics states that the probability $\it{P(n,x)}$ of a single primary electron to 
produce an avalanche with $\it{n}$ electrons, while propagating from the origin to a point $\it{x}$,
 follows an exponential law:

\begin{equation}
\label{eqn2}
\it{P(n,x)} \simeq \frac{1}{\bar{n}(x)}e^{\frac{-n}{\bar{n}(x)}}
\end{equation}

\noindent Here, $\it{\bar{n}}$ is the mean number of avalanche electrons. 
We define $\it{f}$ to be the relative variance of the mean value. 
For an exponential distribution, $\it{f}{\sim1}$ and the photon detection efficiency falls 
exponentially with increasing threshold. 

At higher field values, the avalanche size distribution was found to depart from the monotonically 
falling exponential; it exhibits a rounded peak. 
The probability of ionization by an electron can no longer be considered totally independent of its 
history \cite{Ref30, Ref31}. 
The assumption that all the electrons take part in the multiplication process with equal probability
 must be abandoned. 
The ionization mean-free-path becomes comparable to that for excitation and other inelastic 
processes. 
The charge distribution under these circumstances, introduced by Byrne, is known as Polya 
distribution \cite{Ref32, Ref33}. 
Its derivation assumes that the ionization probability per unit path-length depends on the current 
size of the avalanche through a dimensionless parameter $\theta$. 
The probability of a single primary electron to produce an avalanche with $\it{n}$ electrons, while 
propagating from the origin to a point $\it{x}$, follows a ``peaked'' Polya distribution 
\cite{Ref30, Ref31, Ref32}:

\begin{equation}
\label{eqn3}
\it{P(n,x)} = \left[\frac{(1+\theta)n}{\bar{n}(x)}\right]^{\theta} e^{\left[{\frac{-(1+\theta)n}{\bar{n}(x)}}\right]}
\end{equation}

Relative to an exponential decay, detectors providing single-photoelectron Polya-like spectra are 
expected to have superior detection efficiency for a given electronics threshold. 
For each detector configuration, the relative variance $\it{f}$ and most probable avalanche size
$\it{n_{mp}}$ are given by \cite{Ref34, Ref35}:

\begin{equation}
\label{eqn4}
\it{f} = \frac{1}{1+\theta}
\end{equation}

and

\begin{equation}
\label{eqn5}
\it{n_{mp}} = \frac{\it{\bar{n}}\theta}{1+\theta}
\end{equation}

\noindent respectively. 
Thus, the larger the $\theta$-value, the smaller the variance and the closer the most probable 
charge value is to the mean. 
Namely, for a given charge gain and electronics threshold the larger the $\theta$-value the 
better the single-electron detection efficiency is expected to be. 

The effect of inelastic and ionizing collision on the avalanche size distribution can be understood 
in terms of a simple model discussed in \cite{Ref32}. 
During the electron drift, after each interaction, it either ionizes or loses its kinetic energy by 
other, non-ionizing inelastic collisions. 
At electron energies close to the ionization threshold, the cross-section for ionization is still 
significantly smaller than the sum of the inelastic cross-sections (including excitations). 
With increasing energy, ionization gradually becomes the dominant process; thus, with increasing 
the $\it{E/p}$ - the relative frequency of ionizing collisions is enhanced. 
The shape of the distribution in this case is determined by the ionization yield $\it{Y}$ which is 
defined as:

\begin{equation}
\label{eqn6}
\it{Y} = \frac{\it{N_{ion}}}{\it{N_{ion}+N_{inel}+N_{exc}}}
\end{equation} 

\noindent Here, $\it{N_{ion}}$, $\it{N_{exc}}$ and $\it{N_{inel}}$ are the number of ionizations, 
excitation, and other inelastic collisions, respectively. 

The relative variance $\it{f}$ is, then, related to $\it{Y}$ by:

\begin{equation}
\label{eqn7}
\it{f} \approx \frac{\it{1-Y}}{\it{1+Y}}
\end{equation}

\section{Experimental setup}

In the present work, a single and a double-stage RPWELL-based detector have been investigated. 
The setup consisted of UV- and X-ray sources, a vessel containing the detector elements and a readout 
system.  
The detector vessel was equipped with two windows - a $50~\mu\mathrm{m}$ thick Kapton one for the 
X-rays and a quartz one for the UV photons.

Two radiation sources were used: a self-triggered homemade $\mathrm{H_2}$ discharge lamp emitting 
${\sim}160~\mathrm{nm}$ UV photons and a $^{55}\mathrm{Fe}$ source emitting $5.9~\mathrm{keV}$ 
X-ray photons. 
The $\mathrm{H_2}$-lamp pulse rate was controlled by the voltage supplied to the lamp 
(here, $3.7~\mathrm{kV}$), by a, HK model 6900 Dual MWPC power supply. 
To ensure counting only single-photon events, the UV-flux was attenuated with VUV natural filters 
(Oriel) and a $1~\mathrm{mm}$ diameter collimator. 
The attenuation was set to detect one single UV-photon signal by the detector for every 10 trigger 
signals - resulting in the current setup in single-photon event rate of ${\sim}1~\mathrm{Hz/mm^2}$.

The detector vessel was flushed with either $\mathrm{Ne/CH_{4}}$ ($2\%$, $5\%$, $10\%$ and $15\%$ 
$\mathrm{CH_{4}}$) or with $\mathrm{Ar/5\%CH_{4}}$ to study the effect of gas mixtures and the 
quencher concentration on the single-electron spectrum.

Details of the two, $30\times30~\mathrm{mm^2}$ FR4-made detector configurations are depicted in 
Fig.~\ref{Detector}. 
The single-stage detector is a $0.8~\mathrm{mm}$ thick RPWELL, coated with CsI photocathode 
($\sim300~\mathrm{nm}$ thick) on  its top THGEM-electrode surface; it is preceded by a 
$5~\mathrm{mm}$ drift gap and a drift cathode. 
The double-stage detector has a CsI-coated $0.4~\mathrm{mm}$ thick double-sided THGEM 
pre-amplification element, separated by a $2~\mathrm{mm}$ transfer gap from the 
$0.4~\mathrm{mm}$ thick RPWELL multiplication stage. 
The latter has a single-sided Cu clad at the top (single Cu-THGEM). 
All THGEM electrodes had hexagonal hole layout, hole diameter $0.5~\mathrm{mm}$, hole spacing 
$0.96~\mathrm{mm}$ and hole-rim $0.1~\mathrm{mm}$. 
In both configurations, the RPWELL structure was formed by coupling the THGEM bottom to a 
$0.4~\mathrm{mm}$ thick Semitron ESD225 \cite{Ref36} resistive plate (bulk resistivity of 
$10^9~\Omega \mathrm{cm}$). 
The resistive plate was coupled to a Cu-coated readout anode using 3MTM Electrically Conductive 
Adhesive Transfer Tape 9707.

The transfer electric field, ${E_{Transfer}}$, value in the double-structure, was set to of 
$0.5~\mathrm{kV/cm}$; that of the drift field ${E_{Drift}}$, was set to $0.5~\mathrm{kV/cm}$, to collect the ionization electrons induced by 
X-rays; for an efficient detection of UV-photons, ${E_{Drift}}$ value was set to zero \cite{Ref37}. 
The electrodes were polarized through low-pass filters (LPF), by CAEN N1471H HV power supplies. 
In all configurations, the anode was kept at ground potential and the signals were recorded through 
a charge-sensitive pre-amplifier (CAEN A1422). 
They were further processed through an Ortec 572 A linear amplifier with $2~\mu\mathrm{sec}$ 
shaping time. 
The acquisition was performed either by a Tektronix MSO 5204B Mixed Signal Oscilloscope or by a 
multi-channel analyzer (MCA Amptek 8000A).

In the current study, the measured vacuum QE-value of the CsI photocathode was ${\sim}15\%$ at 
$160~\mathrm{nm}$; no efforts were made to enhance its value. 
Its assembly in the detector vessel was performed under controlled $\mathrm{N_2}$ atmosphere.

\section{Experimental methodology and numerical analysis}

Taking into account potential charging-up effects \cite{Ref38}, some gain-stabilization was 
necessary prior to the measurements. 
Therefore, single-photoelectron charge spectra were measured according to the following protocol: 

\begin{itemize}
\item{Switching ``off'' the source and voltages;} 
\item{Flushing the detector with the selected gas mixture at $50~\mathrm{SCCM}$ for 
$2~\mathrm{hours}$;}  
\item{Switching ``on'' the voltages and the source(s) and waiting for $30~\mathrm{minutes}$ for the 
gain stabilization;} 
\item{Data collection for $30~\mathrm{minutes}$;} 
\item{Switching off the source and measuring the noise for $30~\mathrm{minutes}$;}
\end{itemize}

The signal spectra were obtained by subtracting the measured noise spectra from the measured data. 
In our experimental configuration, the electronic noise was ${\sim}10^4$ electrons.

Each experimental spectrum was fitted to a generic Polya distribution (eqn.~\ref{eqn3}). 
The mean gain value was extracted with its corresponding value for the Polya parameter $\theta$. 
Experimentally, for each detector configuration, the relative single-electron detection efficiency 
was estimated (for a given electronic threshold) as dividing the number of counts at a given 
voltage for a particular configuration by the number of counts at the maximal voltage for the same 
configuration - normalized to the number of lamp triggers, 150,000 per experiment. 
The statistical error in all the measurements is at the level of $1\%$. 
The error bars are too small to be seen in the plots presented below.

In addition, a set of Monte Carlo simulations was carried out using the same analytical function 
(eqn.~\ref{eqn3}) with experimentally obtained detector gain and $\theta$ values as input parameters. 
For 10,000 events, a histogram was filled, representing the expected single-electron spectrum for 
given gain and $\theta$ values. 
The absolute single-electron detection efficiency was estimated by counting events with total 
number of electrons above a given threshold.

It should be worth mentioning here that the reported detector gain values presented in this study 
could vary due to charging-up effects in presence of X-rays \cite{Ref37}. 
By comparing the single-electron spectra before and after X-ray irradiation 
(${\sim}30~\mathrm{Hz/mm^2}$) it was observed that the gain is reduced by a factor of ${\sim}1.4$ 
whereas, the $\theta$ parameter varied by a factor ${\sim}1.2$. 
A detailed study on the charging up effects is beyond the scope of the present work; it will be 
discussed elsewhere.

\section{Results}

Polya-like distributions were recorded with single photoelectrons, in the single and double-stage 
RPWELL-based detectors (shown on Fig.~\ref{Detector}), with different gas mixtures and voltage 
configurations. 
The effects of the latter on the Polya parameters $\theta$ and $\it{f}$ and on the single-electron 
detection efficiency are discussed. 

\subsection{Single- and double-stage RPWELL-based detectors}

\begin{figure}[hbt]
\centering
\subfigure[]
{\label{Geo-Spec}\includegraphics[scale=42]{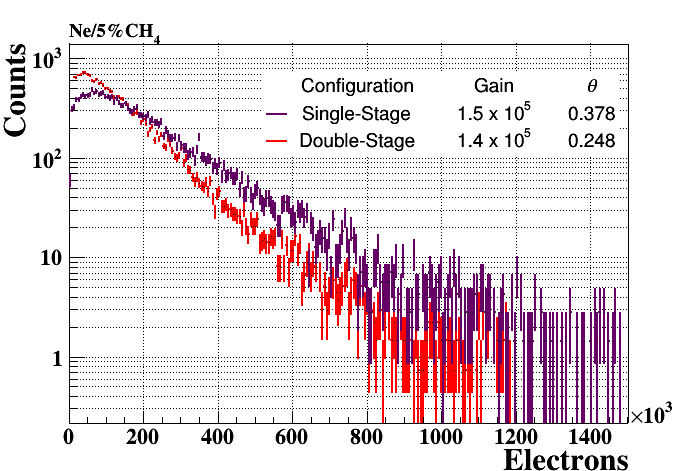}}
\subfigure[]
{\label{Geo-Gain}\includegraphics[scale=0.55]{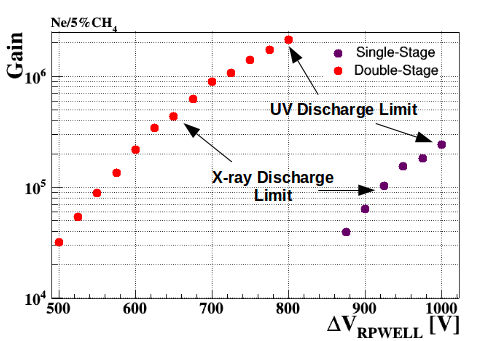}}
\subfigure[]
{\label{Geo-Theta}\includegraphics[scale=42]{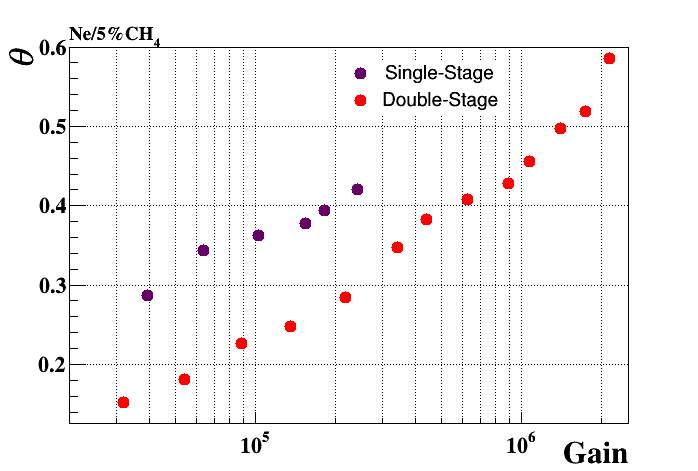}}
\subfigure[]
{\label{Geo-f}\includegraphics[scale=42]{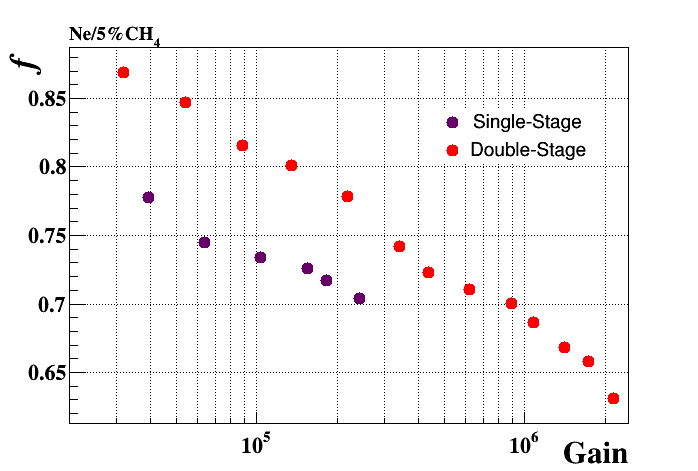}}
\caption{(a) Single-electron spectra recorded in the single and double-stage RPWELL detectors
(of Fig.~\ref{Detector}), at similar charge gains of ${\sim}\it{1.4 \times 10^5}$. (b) Variation of  gain
vs ${\Delta V_{RPWELL}}$, (c) $\theta$ and (d) relative gain variance $\it{f}$ as a function
of gain in $\mathrm{Ne/5\%CH_{4}}$, in the single and double-stage RPWELL-based detectors. For
double-stage configuration ${\Delta V_{THGEM}}~=~700~\mathrm{V}$, ${E_{Transfer}~=~0.5~\mathrm{kV/cm}}$.
${E_{Drift}}~=~0~\mathrm{kV/cm}$. The spectra in (a) are normalized for equal number of counts.}
\label{Geometry}
\end{figure}

UV-induced single-electron spectra were measured with the double-stage and single-stage RPWELL in 
$\mathrm{Ne/5\%CH_{4}}$ at different voltage settings. 
The charge spectra, at a gain of ${\sim}1.4 \times 10^5$, are shown in Fig.~\ref{Geo-Spec}.  
In the double-stage structure, the voltage difference across the THGEM ($\Delta V_{THGEM}$) was 
kept constant at $700~\mathrm{V}$; it was set to maximize the extraction efficiency of the 
photoelectrons and their collection into the holes \cite{Ref38}. 
The RPWELL voltage ($\Delta V_{RPWELL}$), in the double-stage, was set to $575~\mathrm{V}$ to get 
similar total-gain value to that of the single-stage detector (at 
$\Delta V_{RPWELL} = 950~\mathrm{V}$).  

The gain variation with $\Delta V_{RPWELL}$ is shown in Fig.~\ref{Geo-Gain} for the two detector 
configurations. 
Each spectrum was fitted with eqn.~\ref{eqn3}  to yield the values of the gain and $\theta$. 
The double-stage RPWELL reached a maximal gain with UV-photons of ${\sim}2.1 \times 10^6$ 
(${\sim}4 \times 10^5$ with X-rays); in the single-stage detector the gain was limited to ${\sim}2.5 \times 10^5$ - 
both under stable discharge-free operation conditions. 

As expected, the $\theta$-value increases with the gain (Fig.~\ref{Geo-Theta}). 
The relation between $\theta$ and the gain, measured with the two detectors, exhibits the same 
linear trend. 
The comparison with the double-stage detector shows that the $\theta$ value is lower than that of 
the single-stage RPWELL, for equal gain. 
It is due to the charge development and ``saturation'' in a single hole. 
However, as the double-stage detector reaches higher gain values, the maximum achievable $\theta$ 
parameter is higher, reaching values ${\sim}0.6$. 

The relative gain variance ``$\it{f}$'' is plotted as a function of the gain in Fig.~\ref{Geo-f}. 
As discussed earlier, with the increasing gain value (thus, increasing $\theta$ parameter), 
``$\it{f}$'' decreases. 
For a single-stage detector, at a fixed amplification gap, increasing the electric field 
($\Delta V_{RPWELL}$) leads to an increase of the relative frequency of the ionizing collisions; 
therefore, according to eqn.~\ref{eqn6}, resulting in a reduction of the relative variance \cite{Ref32}. 

It was also found that for the same gain value, the relative gain variance is higher for a 
double-stage configuration than for the single-stage one. 
This is expected, since in a double-stage configuration, the electric field per stage is lower than 
in a single-stage one, resulting in a decrease of the ionization probability. 
Also, the loss of electrons while transferred from the pre-amplification stage and the RPWELL 
amplification one gives rise to the additional avalanche-size fluctuations. 

It should be mentioned here that, for the double-stage configuration under X-ray irradiation, the 
maximum applied $\Delta V_{RPWELL}$ was limited to $650~\mathrm{V}$, with a 
$\Delta V_{THGEM} = 700~\mathrm{V}$. 
At these voltage configurations, the maximal achievable gain was ${\sim} 4 \times 10^5$ , with a 
corresponding $\theta$ parameter for single electrons of ${\sim}0.38$ (Fig.~\ref{Geo-Theta}). 

\begin{figure}[hbt]
\centering
\subfigure[]
{\label{Eff-Theory}\includegraphics[scale=42]{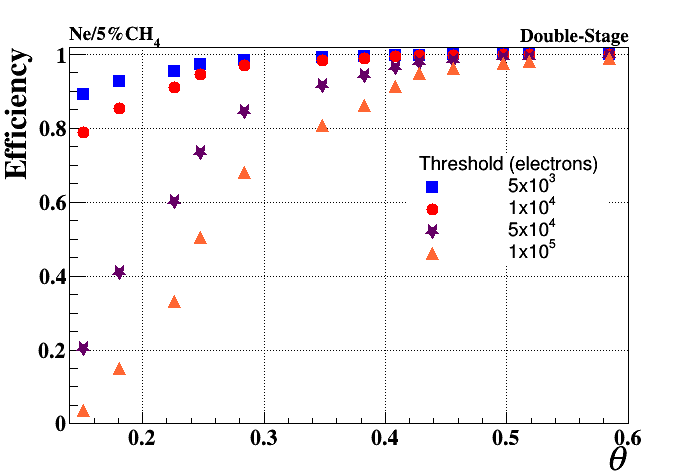}}
\subfigure[]
{\label{Eff-Expt}\includegraphics[scale=42]{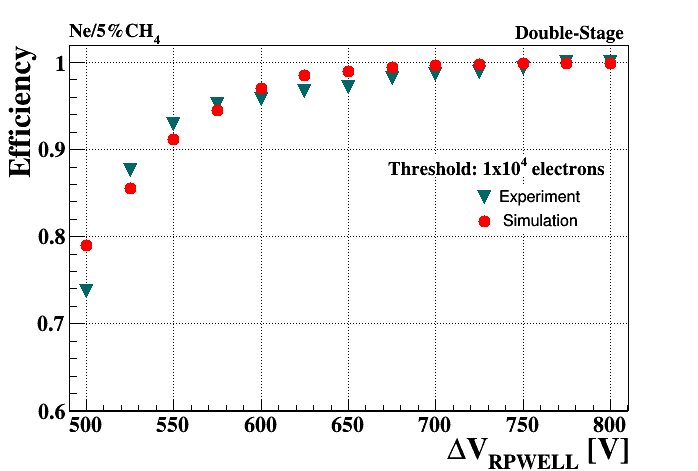}}
\caption{Double-stage RPWELL detector: (a) calculated single-electron efficiency as a function of
$\theta$, for different avalanche-electron thresholds in $\mathrm{Ne/5\%CH_{4}}$. (b) Comparison of
the relative single-electron detection efficiency between experimental and simulation results, as
function of ${\Delta V_{RPWELL}}$. ${\Delta V_{THGEM}}~=~700~\mathrm{V}$, ${E_{Transfer}~=~0.5~\mathrm{kV/cm}}$.
${E_{Drift}}~=~0~\mathrm{kV/cm}$. In both plots, the statistical error bars are too small to be seen.}
\label{Efficiency}
\end{figure}

Numerically-estimated single-electron detection efficiency values are plotted in 
Fig.~\ref{Eff-Theory} as a function of $\theta$ for different avalanche-electrons thresholds. 
Note that even at low $\theta$ values, e.g. ${\sim}0.25$, a high efficiency ($>90\%$) can be obtained 
with a threshold of $10^4$ electrons or less. 
The comparison of the relative efficiency between the experimental estimate and the numerical one 
is presented in Fig.~\ref{Eff-Expt}. The good agreement validates the numerical analysis method.

A comparison of numerically-estimated efficiencies for different gain values in the single and 
double-stage detectors is given in Table~\ref{table-1}. 
The avalanche-electron threshold was fixed to $10^4$ electrons. 
The efficiency of a detector having an exponential distribution, for the same gain values, was 
estimated for comparison. 
It was found that for the same gain value, the single-stage detector provides better efficiency 
than that of the double-stage detector due to the higher $\theta$-value. 
However, with the increase of the gain, the difference reduces. 
Note that the efficiency value for an exponential distribution at the gain of $2.1 \times 10^6$ is a 
hypothetical one. 
It was reached with the double-stage RPWELL, due to its ``charge-quenching'' property; it would be 
hard to reach in other detector configurations.

\begin{table}[]
\centering
\label{table-1}
\begin{tabular}{|c|c|c|c|c|c|}
\hline
\multirow{2}{*}{Gain} & Exponential & \multicolumn{4}{c|}{Polya} \\
\cline{3-6}
& Efficiency & Detector & $\theta$ & Efficiency & Comments \\ 
\hline
\multirow{2}{*}{$6.4 \times 10^4$} & \multirow{2}{*}{$86.3\%$} & Single-Stage & 0.34 & $90.8\%$ & Stable under UV\\
\cline{3-5}
& & Double-Stage & 0.20 & $87.5\%$ & and X-ray\\
\hline
\multirow{2}{*}{$2.4 \times 10^5$} & \multirow{2}{*}{$95.9\%$} & Single-Stage & 0.42 & $98.6\%$ & Stable under UV \\
\cline{3-6}
& & Double-Stage & 0.30 & $97.4\%$ & Stable under UV and X-ray\\
\hline
$2.1 \times 10^6$ & $99.5\%$ & Double-Stage & 0.59 & $99.9\%$ & Stable under UV \\
\hline
\end{tabular}
\caption{Comparison of efficiency in single and double-stage RPWELL-based detectors in
$\mathrm{Ne/5\%CH_4}$. The electron threshold is $10^{4}$ electrons.}
\end{table}

\subsection{Double-stage RPWELL - gain and $\theta$ dependence on THGEM voltage}

We studied the effect of ${\Delta V_{THGEM}}$ on the gain and on the $\theta$ parameter for a fixed
${\Delta V_{RPWELL}}$. The drift and transfer fields were kept at $0$ and $0.5~\mathrm{kV/cm}$, respectively.

\begin{figure}[hbt]
\centering
\subfigure[]
{\label{THGEM-Gain}\includegraphics[scale=42]{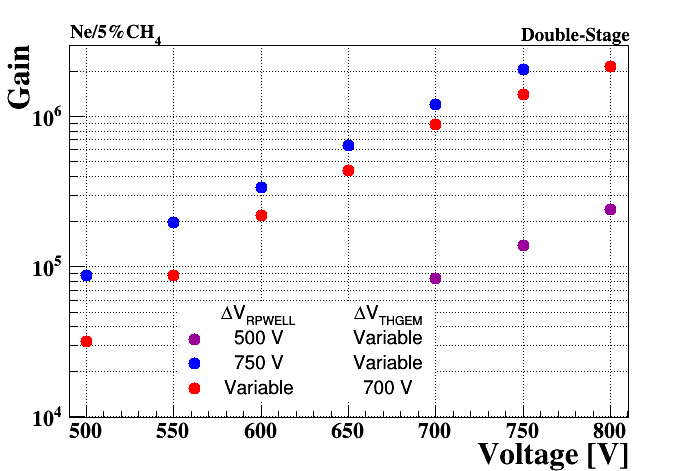}}
\subfigure[]
{\label{THGEM-Theta}\includegraphics[scale=42]{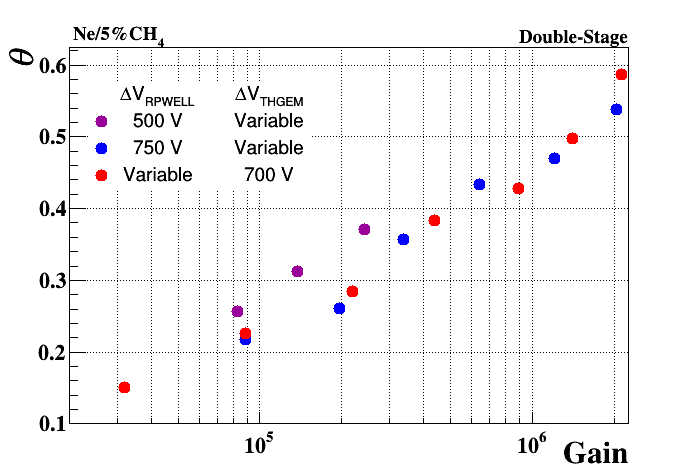}}
\caption{A double-RPWELL configuration in $\mathrm{Ne/5\%CH_{4}}$: (a) gain variation as a function 
of ${\Delta V_{RPWELL}}$ (with  ${\Delta V_{THGEM}}~=~700~\mathrm{V}$) and of ${\Delta V_{THGEM}}$ (with 
${\Delta V_{RPWELL}}~=~750~\mathrm{V}$ and $500~{V}$); (b) corresponding $\theta$ variation as a function 
of the total gain. ${E_{Transfer}~=~0.5~\mathrm{kV/cm}}$ and ${E_{Drift}}~=~0~\mathrm{kV/cm}$.}
\label{THGEM}
\end{figure}

Fig.~\ref{THGEM-Gain} shows the gain dependency on ${\Delta V_{THGEM}}$ for two different values of 
${\Delta V_{RPWELL}}$ and with ${\Delta V_{RPWELL}}$ for a fixed  ${\Delta V_{THGEM}}$.  
${\Delta V_{THGEM}}$ was limited to $800~\mathrm{V}$ before the occurrence of occasional discharges,
 resulting in a maximum gain of $\sim2.8\times10^5$ for ${\Delta V_{RPWELL}}~=~500~\mathrm{V}$. 
For the two other  configurations, with higher applied ${\Delta V_{RPWELL}}$-values, gains up 
to $\sim2.1\times10^6$ were reached under stable conditions. 
The dependency of the $\theta$ parameter on the gain (Fig.~\ref{THGEM-Theta}) is roughly the same 
for the three configurations.

\subsection{Double-stage RPWELL with different gas mixtures}

\subsubsection{Effect of quencher concentration}

\begin{figure}[hbt]
\centering
\subfigure[]
{\label{Quencher-Spec}\includegraphics[scale=42]{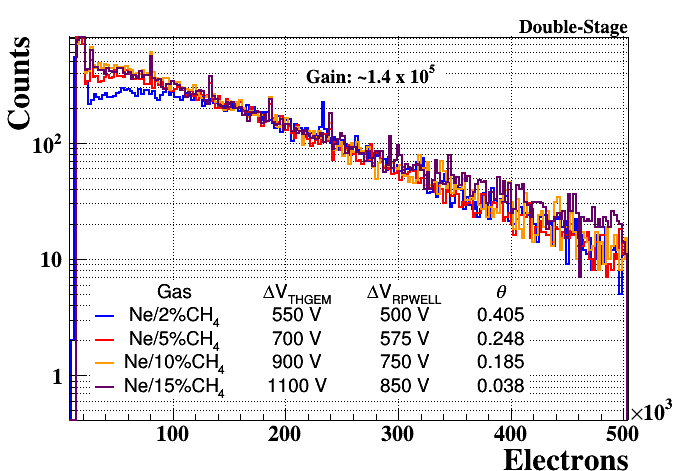}}
\subfigure[]
{\label{Quencher-Gain}\includegraphics[scale=42]{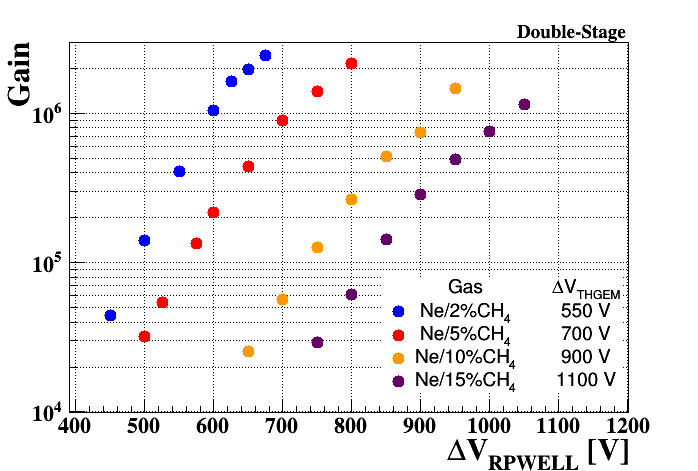}}
\subfigure[]
{\label{Quencher-Theta}\includegraphics[scale=42]{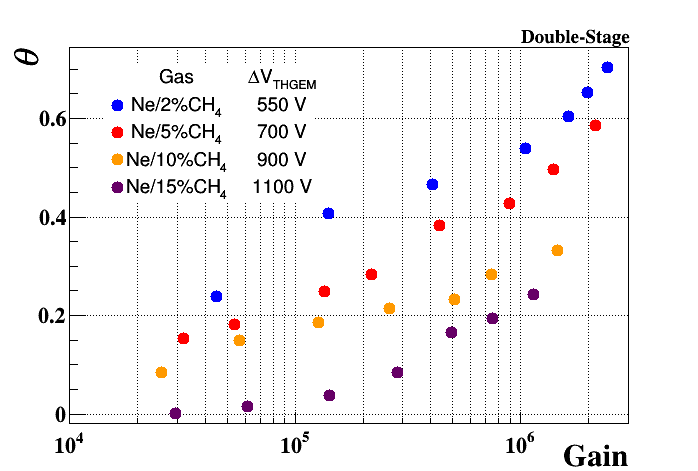}}
\subfigure[]
{\label{Quencher-f}\includegraphics[scale=42]{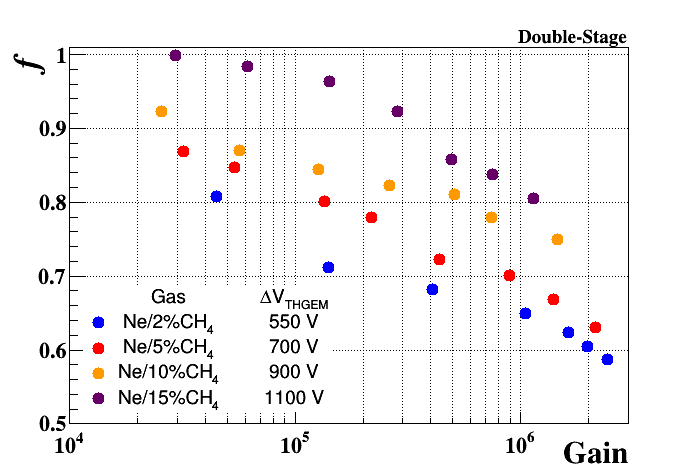}}
\caption{Double-RPWELL configuration in $\mathrm{Ne}$ with $\mathrm{CH_{4}}$-quencher concentrations of $2\%$, $5\%$, $10\%$ and $15\%$: (a) single-electron spectra for a total-gain value for each 
 ${\sim}1.3-1.4 \times 10^{5}$ - reached by adjusting ${\Delta V_{RPWELL}}$ and ${\Delta V_{THGEM}}$
 values; (b) gain variation vs ${\Delta V_{RPWELL}}$; ${\Delta V_{THGEM}}$ adjusted to maximum value
 below discharge onset; (c) corresponding $\theta$ and (d) relative gain variance $\it{f}$ vs total 
gain. ${E_{Transfer}~=~0.5~\mathrm{kV/cm}}$ and ${E_{Drift}}~=~0~\mathrm{kV/cm}$.}
\label{Quencher}
\end{figure}

The dependence of single-electron spectral shape and, consequently, of the $\theta$ parameter, on
the quencher concentration has been studied in $\mathrm{Ne/CH_4}$ mixtures, having $2$, $5$, $10$
and $15\%$ $\mathrm{CH_{4}}$ concentrations.
Fig.~\ref{Quencher-Spec} depicts the single-electron spectra for equal gain of $\sim1.4 \times 10^5$
for the four-quencher concentrations.
One can notice the more pronounced Polya peak at the lower $\mathrm{CH_4}$ concentrations.

Fig.~\ref{Quencher-Gain} shows the trends of the gain with respect to ${\Delta V_{RPWELL}}$ for the
four different concentrations.
In all measurements, ${\Delta V_{THGEM}}$ was set to the highest value allowing for stable, discharge-free
operation to maximize the electron extraction efficiency from the photocathode.
The dependence of $\theta$ parameter and relative variance $\it{f}$-parameter on the gain, at
different quencher concentrations, are shown in Fig.~\ref{Quencher-Theta} and Fig.~\ref{Quencher-f},
 respectively.
In presence of the quencher, the ratio of  ionizing versus other inelastic collisions is shifted in
favor of the other inelastic scattering \cite{Ref32}.
Therefore, the avalanche size distribution broadens up (according to eqn.~\ref{eqn6} and \ref{eqn7}) with the
increase of quencher, thus, the relative variance ``$\it{f}$'' increases.
This in turn, results in higher $\theta$ parameter values for lower quencher concentrations - as
also depicted in Fig.~\ref{Quencher-Theta}.

The single-electron detection efficiency for two different values of gain in the four-quencher
concentration is given in Table~\ref{table-2}.
With an avalanche-electron threshold of $10^4$ electrons, for the lower gain value of ${\sim}6 \times 10^4$,
 the efficiency in $\mathrm{Ne/2\%CH_4}$ is ${\sim}4\%$ higher than that of $\mathrm{Ne/15\%CH_4}$.
This rather small difference is due to the higher $\theta$ value in lower quencher concentrations.
However, with the increase of gain,  the difference between the measured $\theta$ values decreases
and the detector reaches full efficiency for quencher concentrations of $2-15\%$.

\begin{table}[]
\centering
\label{table-2}
\begin{tabular}{|c|c|c|c|c|c|}
\hline
\multirow{2}{*}{Gain} & Exponential & \multicolumn{4}{c|}{Polya} \\
\cline{3-6}
& Efficiency & $\mathrm{CH_4}$ Concentration & $\theta$ & Efficiency & Comments \\
\hline
\multirow{4}{*}{$6.1 \times 10^4$} & \multirow{4}{*}{$84.9\%$} & $2\%$ & 0.28 & $89.5\%$ & Stable operation\\
\cline{3-5}
& & $5\%$ & 0.19 & $88.2\%$ & of double-stage\\
\cline{3-5}
& & $10\%$ & 0.16 & $87.9\%$ & under UV and\\
\cline{3-5}
& & $15\%$ & 0.02 & $85.5\%$ & of X-ray\\
\hline
\multirow{4}{*}{$4.9 \times 10^5$} & \multirow{4}{*}{$98\%$} & $2\%$ & 0.45 & $99\%$ & Stable operation\\
\cline{3-5}
& & $5\%$ & 0.41 & $99\%$ & of double-stage\\
\cline{3-5}
& & $10\%$ & 0.25 & $99\%$ & under UV \\
\cline{3-5}
& & $15\%$ & 0.17 & $99\%$ & \\
\hline
\end{tabular}
\caption{Comparison of efficiency as a function of quencher amount in double-stage RPWELL-based detectors in
$\mathrm{Ne/CH_{4}}$. The electron threshold is $10^4$ electrons.}
\end{table}

\subsubsection{Effects of the carrier gas}

\begin{figure}[hbt]
\centering
\subfigure[]
{\label{Primary-Spec}\includegraphics[scale=42]{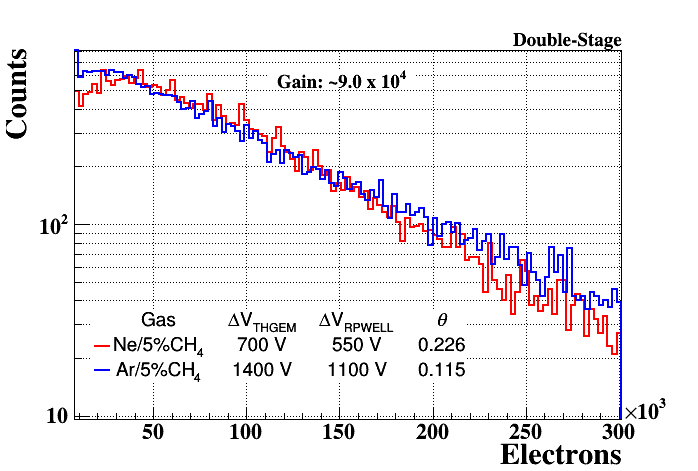}}
\subfigure[]
{\label{Primary-Gain}\includegraphics[scale=42]{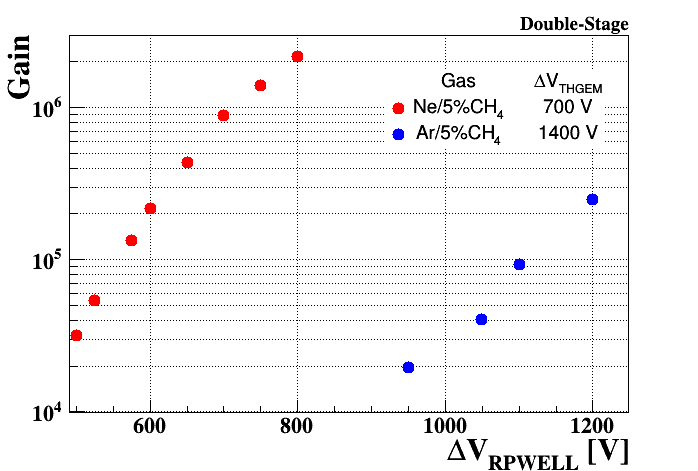}}
\subfigure[]
{\label{Primary-Theta}\includegraphics[scale=42]{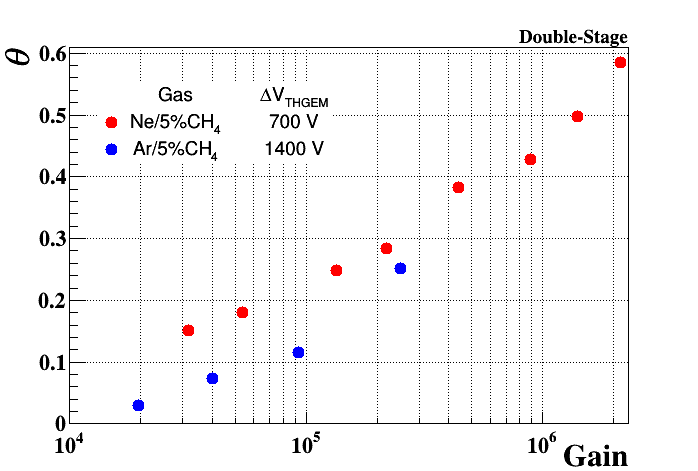}}
\subfigure[]
{\label{Primary-f}\includegraphics[scale=42]{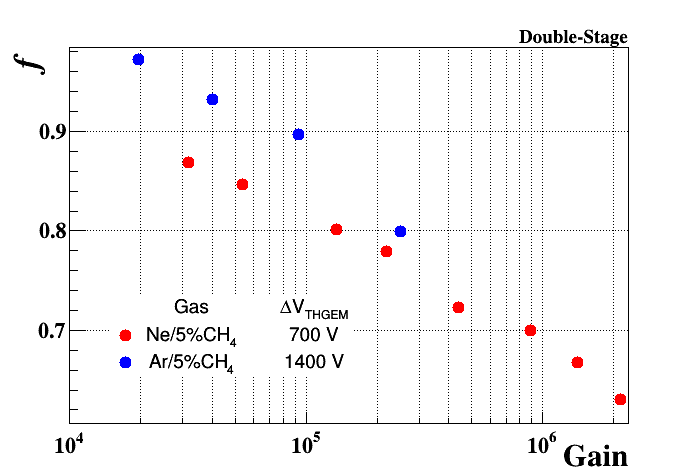}}
\caption{A double-RPWELL configuration in $\mathrm{Ne/5\%CH_{4}}$ and $\mathrm{Ar/5\%CH_{4}}$: (a) 
single-electron spectra at equal total charge gain of ${\sim}10^{5}$, adjusted by the 
${\Delta V_{RPWELL}}$ and ${\Delta V_{THGEM}}$ values, (b) gain variation vs ${\Delta V_{RPWELL}}$; 
(c) corresponding $\theta$ and (d) relative gain variance $\it{f}$. ${E_{Transfer}~=~0.5~\mathrm{kV/cm}}$ 
and ${E_{Drift}}~=~0~\mathrm{kV/cm}$.}
\label{Primary}
\end{figure}

The dependence of the $\theta$ parameter on the carrier gas, for a fixed $\mathrm{CH_{4}}$ quencher
concentration were studied.
Fig.~\ref{Primary-Spec} shows single-electron spectra in $\mathrm{Ne/5\%CH_4}$ and
$\mathrm{Ar/5\%CH_4}$, for similar gain values of ${\sim}10^5$, adjusted by setting different
${\Delta V_{THGEM}}$ values.
The gain curves as function of ${\Delta V_{RPWELL}}$ are shown in Fig.~\ref{Primary-Gain}.
The operation in $\mathrm{Ar/5\%CH_4}$ required higher ${\Delta V_{THGEM}}$ and
${\Delta V_{RPWELL}}$ values.
The maximum achievable gain is higher in the Ne-based mixture, by at least an order of magnitude,
as previously observed in \cite{Ref21, Ref38}.
The dependence of $\theta$ and $\it{f}$ on the gain in these two gas mixtures are shown in
Fig.~\ref{Primary-Theta} and Fig.~\ref{Primary-f}, respectively.
In $\mathrm{Ne/5\%CH_4}$, the $\it{f}$  value is slightly lower than in $\mathrm{Ar/5\%CH_4}$; it
agrees with measurements in Micromegas detectors \cite{Ref32, Ref34}.
In Ar-based mixtures, due to the lower threshold for excitation and larger inelastic cross-section,
the ionization yield is lower, thus enhancing avalanche fluctuations.

\begin{table}[]
\centering
\label{table-3}
\begin{tabular}{|c|c|c|c|c|c|}
\hline
\multirow{2}{*}{Gain} & Exponential & \multicolumn{4}{c|}{Polya} \\
\cline{3-6}
& Efficiency & Gas & $\theta$ & Efficiency & Comments \\
\hline
\multirow{2}{*}{$4.0 \times 10^4$} & \multirow{2}{*}{$78\%$} & $\mathrm{Ne/5\%CH_4}$ & 0.16 & $81\%$ & Stable operation of double-stage\\
\cline{3-5}
& & $\mathrm{Ar/5\%CH_4}$ & 0.07 & $79\%$ & under UV and X-ray\\
\hline
\multirow{2}{*}{$2.5 \times 10^5$} & \multirow{2}{*}{$96\%$} & $\mathrm{Ne/5\%CH_4}$ & 0.30 & $98\%$ & Stable operation of \\
\cline{3-5}
& & $\mathrm{Ar/5\%CH_4}$ & 0.25 & $97\%$ & double-stage under UV \\
\hline
\end{tabular}
\caption{Comparison of efficiency as a function of carrier gas in double-stage RPWELL-based detectors. 
The electron threshold is $10^4$ electrons.}
\end{table}

The numerically estimated single-electron detection efficiency for two different values of gain in 
$\mathrm{Ne/5\%CH_4}$ and $\mathrm{Ar/5\%CH_4}$ is given in Table~\ref{table-3}. 
For the lower gain value of ${\sim}4 \times 10^4$, the efficiency in $\mathrm{Ne/5\%CH_4}$ is slightly 
higher than in $\mathrm{Ar/5\%CH_4}$ due to the higher $\theta$ value. 
However, with the increase of gain, the difference between the efficiencies is smaller.

\section{Summary and discussion}

In this work, we investigated single-electron spectra obtained with single-stage and double-stage 
RPWELL-based detectors. 
The goal was to evaluate the potential advantage of their operation in conditions yielding 
Polya-like single distributions; the latter result from charge-avalanche saturation in the RPWELL 
holes. 
The detectors' response was single-photoelectrons emitted from a CsI photocathode deposited on the 
multiplier's surface. 
Their performance was  studied in various gas mixtures and electric-field settings.

Operating in $\mathrm{Ne/5\%CH_4}$, the single-stage RPWELL detector reached gains of 
${\sim}2.5 \times 10^5$ in a discharge-free mode; the single-electron Polya-like distribution had a 
$\theta$-value of ${\sim}0.36$. 
The double-stage THGEM/RPWELL detector reached gains of ${\sim}2.1 \times 10^6$ in stable conditions; the 
Polya-like distribution reached a $\theta$-value of ${\sim}0.6$.

Similar trends are shown for the variation of the Polya parameters $\theta$ and $\it{f}$ as a 
function of gain in both RPWELL detector configurations. 
Note, however, that for equal gain, the single-stage detector yielded more pronounced Polya-like 
distributions, with higher $\theta$-value and thus narrower relative gain variance $\it{f}$. 
This is expected since in a double-stage configuration, the electric field per stage is lower than 
in a single-stage one, resulting in a decrease of the ionization probability (gain fluctuations) 
with respect to other inelastic processes. 
In addition, avalanche-electron losses between the two stages result in additional avalanche-size 
fluctuations.

The experimental values of $\theta$ and gain, followed by numerical calculations, yielded the 
single-electron detection-efficiency values; they are presented as function of the threshold 
(number of avalanche-electrons). 
For a given threshold of $10^4$ electrons, the numerical estimation suggests that for a same gain 
of ${\sim}6 \times 10^4$, the efficiency in the single-stage detector (${\sim}91\%$) is superior to that of 
the double-stage one (${\sim}87.5\%$), due to the higher $\theta$ value. 
Also, for the single-stage detector, relative to an exponential distribution at the same gain, the 
expected single-electron detection efficiency is about ${\sim}6\%$ higher ($91\%$ relative to $86\%$).
 However, with the increase of the gain, for the same threshold, both detectors attain similar 
efficiency values. 
The double-stage detector yielded an efficiency of ${\sim}97\%$ at a gain of ${\sim}2 \times 10^5$, with 
$\theta = 0.3$. 
In the presence of 5.9 keV X-ray, the maximal achievable stable gain is limited to ${\sim}4 \times 10^5$ , 
with a $\theta$ value of ${\sim}0.38$ and $99\%$ detection efficiency.

The amount of quencher added to the carrier gas has a major role in reducing instabilities, which 
are mainly due to photon-induced secondary avalanches. 
It also affects the electron-transport parameters and enhances the extraction efficiency of 
photoelectrons from the photocathode in a single-photon detector (the matter is out of the scope of 
this work). 
We studied the effect of the $\mathrm{CH_4}$ quencher concentration on the detector properties, 
gain, $\theta$ and the relative gain variance $\it{f}$. 
It was observed that with increasing quencher concentration, $\theta$-value decreased, and $\it{f}$ 
increased - for a given gain value. 
With an electron threshold of $10^4$, for a gain of $6 \times 10^4$, the $2\%$ $\mathrm{CH_4}$ 
concentration yielded higher efficiency ($89.5\%$) than that of $15\%$ $\mathrm{CH_4}$ 
($85.5\%$ efficiency) due to a lower $\theta$ value in the latter case. 
But, with the increase of gain, the quencher concentration did not affect the efficiency 
significantly, for the same threshold. 
Relative to an exponential distribution at a similar gain of ${\sim}6 \times 10^4$, the single-electron 
efficiency improved by ${\sim}5\%$ for $2\%$ quencher concentration and by ${\sim}1\%$ for $15\%$ 
quencher concentration.

The dependence of $\theta$ and $\it{f}$ on the carrier gas was evaluated for a given 
$\mathrm{CH_4}$ quencher concentration. Significant avalanche fluctuations were observed in 
$\mathrm{Ar/5\%CH_4}$; it is due to low ionization yield resulting in higher relative gain variance. 
$\mathrm{Ne/5\%CH_4}$, with lower avalanche fluctuations, showed a higher breakdown limit - thus a 
lower discharge probability. 
Therefore, $\mathrm{Ne/CH_4}$ proved to be superior to $\mathrm{Ar/CH_4}$ in detecting 
single-photoelectrons. 
The numerical study suggested that depending on the electron threshold, at lower gains, the 
detector operation in $\mathrm{Ne/CH_4}$ yields higher efficiency than in $\mathrm{Ar/CH_4}$; with 
increasing gain, both mixtures reach similar efficiency values, e.g. ${\sim}98\%$ at $2.5 \times 10^5$.  
Relative to an exponential distribution at lower gain, the single-electron efficiency improved by 
${\sim}1\%$ for $\mathrm{Ar/5\%CH_4}$ and by ${\sim}4\%$ for $\mathrm{Ne/5\%CH_4}$.

\acknowledgments

This research was supported in part by the Nella and Leon Benoziyo Center for 
High Energy Physics at the Weizmann Institute of Science and Grant No 
713563 from the Israeli Science Foundation (ISF). 
This work is supported by Sir Charles Clore Prize. Special thanks to Martin 
Kushner Schnur for supporting this research.

\end{document}